# Evaluation of excitation schemes for indirect detection of $^{14}$N *via* solid-state HMQC NMR experiments

Andrew G.M. Rankin,[1*] Julien Trébosc,[1,2] Piotr Paluch,[1,3] Olivier Lafon,[1,4] Jean-Paul Amoureux[1,5*]

[1] Univ. Lille, CNRS, Centrale Lille, ENSCL, Univ. Artois, UMR 8181 – UCCS – Unit of Catalysis and Chemistry of Solids, F-59000 Lille, France
[2] Univ. Lille, CNRS-FR2638, Fédération Chevreul, F-59000 Lille, France
[3] Centre of Molecular and Macromolecular Studies, Polish Academy of Sciences, Sienkiewicza 112, PL-90363 Lodz, Poland.
[4] Institut Universitaire de France, 1 rue Descartes, F-75231 Paris Cedex 05, France.
[5] Bruker Biospin, 34 rue de l'industrie, F-67166 Wissembourg, France.

* Corresponding authors: jean-paul.amoureux@univ-lille.fr    andrew.rankin@univ-lille.fr

**Abstract.** It has previously been shown that $^{14}$N NMR spectra can be reliably obtained through indirect detection *via* HMQC experiments. This method exploits the transfer of coherence between single- (SQ) or double-quantum (DQ) $^{14}$N coherences, and SQ coherences of a suitable spin-1/2 'spy' nucleus, *e.g.*, $^{1}$H. It must be noted that SQ-SQ methods require a carefully optimized setup to minimize the broadening related to the first-order quadrupole interaction (*i.e.*, an extremely well-adjusted magic angle and a highly stable spinning speed), whereas DQ-SQ ones do not. In this work, the efficiencies of four $^{14}$N excitation schemes (DANTE, XiX, Hard Pulse (HP), and Selective Long Pulse (SLP)) are compared using *J*-HMQC based numerical simulations and either SQ-SQ or DQ-SQ $^{1}$H-{$^{14}$N} *D*-HMQC experiments on L-histidine HCl and N-acetyl-L-valine at 18.8 T and 62.5 kHz MAS. The results demonstrate that both DANTE and SLP provide a more efficient $^{14}$N excitation profile than XiX and HP. Furthermore, it is shown that the SLP scheme: (i) is efficient over a large range of quadrupole interaction, (ii) is highly robust to offset and rf-pulse length and amplitude, and (iii) is very simple to set up. These factors make SLP ideally suited to widespread, non-specialist use in solid-state NMR analyses of nitrogen-containing materials.

## 1. Introduction

Nitrogen atoms are ubiquitous in biomolecules, *e.g.*, DNA and proteins, and organic, hybrid and inorganic materials, such as pharmaceuticals, polyamides, functionalized silica, metal-organic frameworks or nitrides. However, the NMR observation of the two nitrogen isotopes, $^{15}$N and $^{14}$N, remains challenging owing to their unfavorable NMR properties.

### 1.1. $^{15}$N

It is a spin-1/2 nucleus, which often leads to high-resolution spectra, even for solids. However, owing to its very low natural abundance (NA = 0.36 %) and its low gyromagnetic ratio, $\gamma_{15N} \approx 0.1\gamma_{1H}$, a major limitation of $^{15}$N NMR is often the lack of sensitivity. This issue can be circumvented by $^{15}$N isotopic enrichment, but this approach can be costly, time-consuming and difficult to achieve. Alternative methods to enhance the NMR sensitivity and detect the $^{15}$N signals in natural abundance consist in the use of (i) indirect detection *via* protons at high Magic Angle Spinning (MAS) frequencies using through-space or through-bond $^{1}$H→$^{15}$N polarization transfers [1,2] or (ii) dynamic nuclear polarization enhanced $^{1}$H → $^{15}$N cross-polarization under MAS (CPMAS) experiments [2,3].

### 1.2. $^{14}$N

Conversely, the $^{14}$N isotope has two major advantages: (i) a very high natural abundance (NA = 99.64 %), and (ii) the fact that the knowledge of the quadrupole interaction of this spin-1 nucleus provides additional information about the local environment of nitrogen atoms.

**Direct detection.**

The direct NMR detection of $^{14}$N nuclei is challenging for several reasons [4–6]. First, it has a spin-1 value and hence no central transition between energy levels $m_N = \pm 1/2$. For a powdered sample, the signal of the two single-quantum transitions, $^{14}$N$^{SQ}$, between energy levels $m_N = 0$ and $\pm 1$ is broadened by the first-order quadrupole interaction (H$_{Q1}$) over a width of $3C_Q/2$, where $C_Q$ is the quadrupolar coupling constant. For this isotope, a value of $C_Q = 7$ MHz has been reported [7], which corresponds to a powder pattern width of 10.5 MHz. This large line broadening severely decreases the sensitivity since the total integrated intensity is spread over a broad spectral width. Second, the sensitivity of directly-detected spectra being proportional to $\gamma^{3/2}$ that of $^{14}$N isotope is further decreased with respect to $^{15}$N by its low gyromagnetic ratio with $\gamma_{14N} \approx 0.072\gamma_{1H}$. Third, the pulse dead-times, which are inversely proportional to $\gamma$, are thus long for $^{14}$N nuclei, whereas the broad linewidths result in short Free-Induction Decays (FID). This combination also decreases the sensitivity of $^{14}$N direct detection. For $^{15}$N isotope, this issue can be circumvented by the use of a spin-echo sequence or the CPMG (Carr-Purcell Meiboom-Gill) variant, which improves the sensitivity by the acquisition of a train of echoes in every scan [8]. However, these methods are not easy to use for the $^{14}$N isotope, because π-pulses are difficult to implement for this spin-1 nucleus. Forth, an additional difficulty for the direct $^{14}$N detection is the need to excite a broad spectral width, which can exceed the excitation bandwidth of the radio-frequency (rf) pulses and the detection bandwidth of the probe. This limitation is exacerbated for $^{14}$N isotope because the rf-fields are proportional to $\gamma$, which is small for $^{14}$N. When the width of the $^{14}$N$^{SQ}$ spectrum exceeds the detection bandwidth of the probe or the excitation bandwidth of the pulses, sub-spectra with different frequency offsets are acquired and then co-added to generate the full wide powder pattern [8]. This piecewise acquisition technique is called VOCS (Variable Offset Cumulative Spectrum). It has been shown that the excitation bandwidth of directly detected $^{14}$N experiments can be increased using adiabatic pulses, such as WURST (wideband uniform-rate smooth-truncation) under static conditions [9] or trains of short rotor-synchronized rf-pulses in the manner of DANTE (Delays Alternating with Nutation for Tailored Excitation) under MAS conditions [10–12]. The sensitivity of directly detected $^{14}$N experiments under static conditions can be further enhanced by an initial polarization transfer from protons using BRAIN-CP (broadband adiabatic inversion cross-polarization) [13].

Nevertheless, a major limitation of the direct $^{14}$N$^{SQ}$ detection under static conditions is the lack of resolution since this signal is broadened by H$_{Q1}$ and the powder patterns of distinct nitrogen sites extensively overlap. Hence, this approach is only suitable for samples with one or two nitrogen sites. MAS can improve the resolution of $^{14}$N$^{SQ}$ spectra, but these experiments require a very precise adjustment and a high stability of the MAS frequency and angle [14–17]. The resolution can be improved by the detection of the $^{14}$N double-quantum ($^{14}$N$^{DQ}$) coherences between energy levels $m_N = \pm 1$ since these coherences are only broadened by the second-order quadrupole interaction (H$_{Q2}$). These coherences can be directly excited and acquired at twice the $^{14}$N Larmor frequency with the overtone sequence. This type of direct excitation has been reported first under static conditions [18,19] and more recently under MAS [4,20,21]. The detection of this usually "forbidden" DQ transition is allowed by a perturbation of the Zeeman spin states by the large quadrupole interaction. The sensitivity of overtone experiments can be enhanced by $^1$H → $^{14}$N CPMAS transfer and DNP [22,23]. However, the overtone $^{14}$N direct excitation and detection experiments are usually less sensitive than the direct detection of $^{14}N^{SQ}$ transitions. Furthermore, the slow nutation of the overtone signal results in a small excitation bandwidth, which may prevent the detection of distinct nitrogen sites. WURST pulses can improve the overtone excitation bandwidth, but they decrease the sensitivity [24].

**Indirect detection.**

For protonated samples, the lack of resolution and sensitivity of direct $^{14}N$ detection can be partially overcome by the indirect detection of $^{14}N^{SQ}$ or $^{14}N^{DQ}$ transitions *via* protons, denoted $^{1}$H-$\{^{14}N^{SQ}\}$ and $^{1}$H-$\{^{14}N^{DQ}\}$ respectively [5,25–27], using two-dimensional (2D) experiments at high MAS frequencies, such as HMQC (heteronuclear multiple-quantum correlation) [25,28], HSQC (heteronuclear single-quantum correlation) [29,30], double cross-polarization [31], or sequences using $^{14}N$ rf-irradiation lasting hundreds of microseconds to generate $^{1}$H-$^{14}N$ multiple-quantum coherences [32]. It has been shown that the HSQC sequence is less sensitive than the HMQC one [29,33]. Furthermore, for HMQC experiments, coherence transfers *via* $^{1}$H-$^{14}N$ dipolar interactions, which are reintroduced by the application of heteronuclear dipolar recoupling schemes, such as $SR4_1^2$ [34], in dipolar-mediated HMQC schemes, are usually faster and more efficient than coherence transfers *via* a combination of $J$-coupling and second-order quadrupolar-dipolar cross-terms, also known as residual dipolar splitting, during the free evolution periods of the conventional $J$-mediated HMQC ($J$-HMQC) scheme [28,35]. For $^{1}$H-$\{^{14}N^{SQ}\}$ HMQC experiments, the broadening of the $^{14}N^{SQ}$ transitions due to the first-order quadrupole interaction is refocused by the rotor synchronization of the indirect evolution period, *i.e.*, $t_1 = nT_R$.

As for the $^{14}N$ direct detection, a challenge for the indirect detection is the creation of $^{14}N$ coherences because the magnitude of $H_{Q1}$ exceeds that of the rf-field by several orders of magnitude. In $^{1}$H-$\{^{14}N\}$ HMQC experiments, $^{14}N$ SQ and DQ coherences have been created and reconverted by: short 'hard' pulses (HP) using high rf-power [36,37], sideband selective long pulses (SLP) [25,28,35,38,39], and DANTE trains [10,40]. Recently, single-sideband XiX (x-inverse x) pulse trains have been introduced in the $^{1}$H-$\{^{14}N^{DQ}\}$ HMQC sequence [41]. These four excitation schemes (HP, SLP, DANTE and XiX) have all been applied at the $^{14}N$ Larmor frequency. $^{1}$H-$\{^{14}N^{DQ}\}$ $D$-HMQC 2D spectra have also been recorded using overtone irradiation at twice the $^{14}N$ Larmor frequency, in the manner of rectangular, WURST and composite pulses [42–44]. Nevertheless, it has been reported that $^{1}$H-$\{^{14}N^{DQ}\}$ $D$-HMQC experiments on amino-acids using rectangular pulses or DANTE trains at the $^{14}N$ Larmor frequency, or overtone pulses are less sensitive than the $^{1}$H-$\{^{14}N^{SQ}\}$ $D$-HMQC ones using rectangular pulses or DANTE trains [35,39,45]. Furthermore, for the γ polymorph of glycine, we have shown that the $^{1}$H-$\{^{14}N^{SQ}\}$ $D$-HMQC experiment using DANTE is slightly more sensitive, but less robust to offset, than its SLP variant [39,40]. The performances of $^{1}$H-$\{^{14}N^{DQ}\}$ $D$-HMQC using XiX excitation have not been compared to those of the other existing $^{1}$H-$\{^{14}N\}$ $D$-HMQC schemes.

Here, we compare the performances of $^{1}$H-$\{^{14}N\}$ $D$-HMQC experiments using HP, SLP, DANTE and XiX schemes, all at the $^{14}N$ Larmor frequency. These sequences are first described using an average Hamiltonian theory. We then analyze the efficiency of these experiments and their robustness to offset, rf-field inhomogeneity and quadrupole interaction. The performances of $^{1}$H-$\{^{14}N\}$ $D$-HMQC schemes using HP, SLP and DANTE schemes are also assessed by experiments on L-histidine.HCl and N-acetyl-L-valine, denoted His and NAV hereafter, respectively.

## 2.    Pulse sequences and theory

Fig.**1** shows the $^1$H-{$^{14}$N} $J$- and $D$-HMQC sequences using rectangular pulses (HP or SLP), the basic $D_1^K$ DANTE, and the XiX schemes for the excitation and the reconversion of $^{14}$N coherences evolving during the $t_1$ period. In the $J$-HMQC sequence, the defocusing and refocusing delays, denoted $\tau_J$, are free evolution periods. In the $D$-HMQC sequence, the SR4$_1^2$ recoupling was applied during the defocusing and refocusing delays, $\tau_D$, in order to reintroduce the $^1$H-$^{14}$N dipolar couplings. During SR4$_1^2$, the $^1$H rf-field is equal to twice the MAS frequency: $\nu_1 = 2\nu_R$. This recoupling reintroduces the space component $|m| = 1$ of the $^1$H-$^{14}$N dipolar couplings and $^1$H chemical shift anisotropy (CSA), whereas it suppresses the $^1$H-$^1$H dipolar couplings, the $^1$H isotropic chemical shifts and the hetero-nuclear $J$-couplings with protons. As the SR4$_1^2$ scheme is non-γ-encoded, the beginning of the two SR4$_1^2$ periods in the $D$-HMQC sequence must be separated by an integer number of rotor periods.

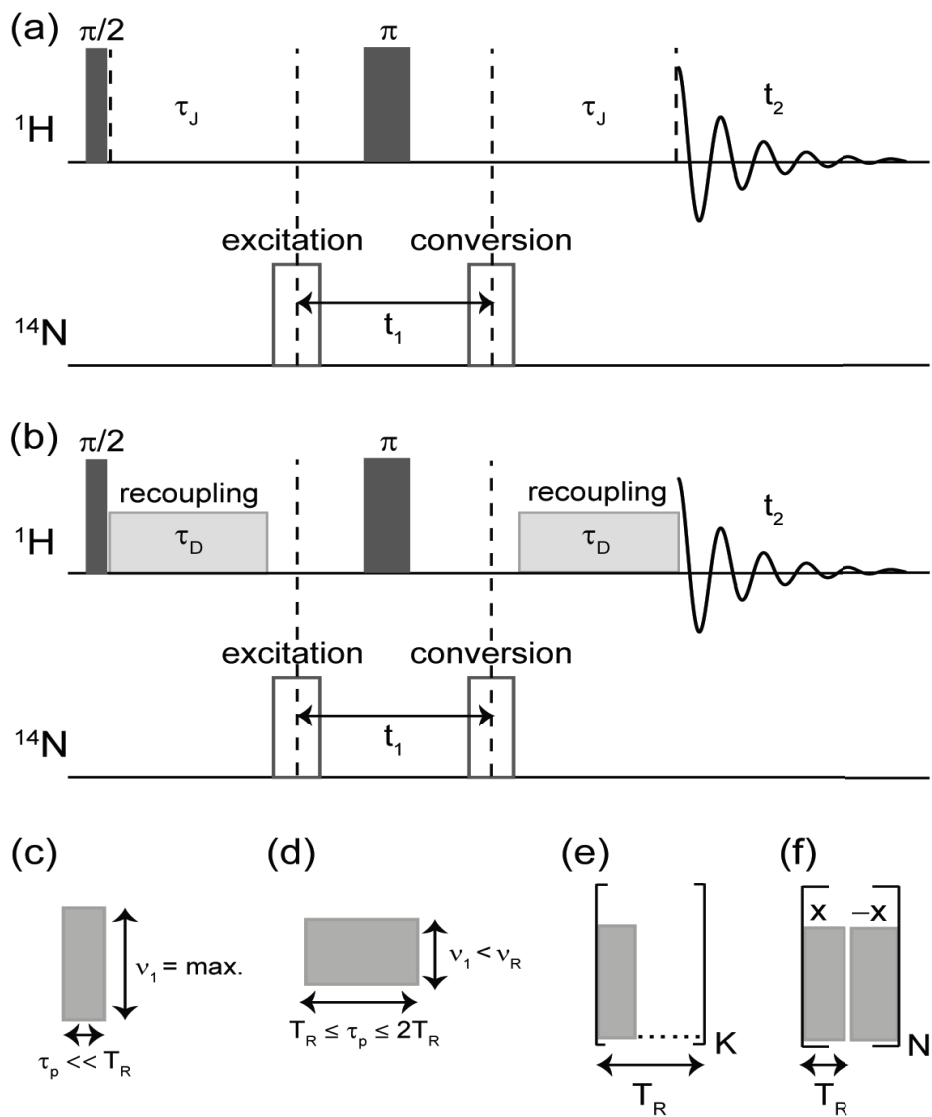

**Fig.1.** $^1$H-{$^{14}$N} (a) $J$- and (b) $D$-HMQC sequences using for the excitation and reconversion of $^{14}$N coherences in place of the open rectangles in (b) on the $^{14}$N channel: (c) HP, (d) SLP, (e) $D_1^K$ and (f) XiX schemes. The $t_1$ period is defined as the delay between the middles of the excitation and reconversion blocks.

For $^1$H-{$^{14}$N} HMQC experiments, the $^1$H $x$-magnetization created by the first pulse evolves during the defocusing period into [40,46]

$$S_x \left(\frac{1+2\cos(2\omega_{HN}\tau)}{3}\right) + S_x Q_z \left(\frac{\cos(2\omega_{HN}\tau)-1}{3}\right) + S_y I_z \sin(2\omega_{HN}\tau) \qquad (1)$$

where $S_x Q_z$ is the $x$-magnetization of proton, leading to a triplet with intensities 1:-2:1 for the coupling with $^{14}N$ nucleus, $S_y I_z$ is the $y$-magnetization of proton, leading to a triplet with intensities 1:0:-1. The angular frequency $\omega_{HN}$ is equal to $\pi J_{HN}$ for $J$-HMQC and to $\omega_{D,HN}$ for $D$-HMQC:

$$\omega_{D,HN} = \frac{1}{4} b_{HN} \sin^2(\beta_{PR}^{D,HN}) \cos(2\varphi). \qquad (2)$$

with

$$\varphi = \gamma_{PR}^{D,HN} + \alpha_{RL}^0 - \omega_R t^0 \qquad (3)$$

In Eqs.2 and 3, $b_{HN}$ denotes the dipolar coupling constant in rad.s$^{-1}$, the Euler angles $\{0, \beta_{PR}^{D,HN}, \gamma_{PR}^{D,HN}\}$ relate the inter-nuclear $^1H$-$^{14}N$ direction to the MAS rotor-fixed frame, and $t^0$ refers to the starting time of the symmetry-based scheme. The subsequent evolution of the density matrix depends on the used excitation scheme.

## 2.1. HP

HP can achieve the uniform excitation of $^{14}N$ powder patterns provided its length $\tau_p$ is sufficiently short [39,40]

$$\tau_p < 2/(3\pi C_Q). \qquad (4)$$

For $C_Q = 1$ MHz, Eq.4 indicates that the uniform excitation requires $\tau_p < 0.2$ μs. The maximal $^{14}N$ rf-field compatible with the specifications of a 1.3 mm double-resonance MAS probe is $\nu_1 \approx 100$ kHz, which results in a flip angle of $\theta < 2\pi\nu_1\tau_p \approx 8°$. This maximum value is much smaller than the optimal angles, $\theta \approx 54.4$ and $90°$, yielding maximal transfer efficiency for $^1H$-$\{^{14}N^{SQ}\}$ and $^1H$-$\{^{14}N^{DQ}\}$ $D$-HMQC experiments, respectively [40]. These small tilt angles decrease the efficiency of $^1H$-$\{^{14}N\}$ HMQC experiments using HP. When Eq.4 is satisfied, the evolution under $H_{Q1}$ during $\tau_p$ is negligible and the Hamiltonian in the jolting-frame during HP is equal to [39,41]

$$\tilde{H} = 2\pi\nu_1 I_x \qquad (5)$$

In Eq.1, a HP corresponding to this Hamiltonian converts $I_z$ into $I_y$ and the quadrupolar order operator $Q_z$ into $K_y$ and $D_x$ operators, which represent respectively the $^{14}N^{SQ}$ coherences along y-axis antiphase with respect to $H_{Q1}$ and the $^{14}N^{DQ}$ coherences along the x-axis. After the first HP, only the signal resulting from either $S_x K_y$ and $S_y I_y$ or $S_x D_x$ operator is detected, according to $^1H$-$\{^{14}N^{SQ}\}$ or $^1H$-$\{^{14}N^{DQ}\}$ HMQC experiment is used, respectively. The coherences evolve under isotropic shift during $t_1$. For $^1H$-$\{^{14}N^{SQ}\}$, the excitation and reconversion HPs must be rotor-synchronized in order to refocus the evolution under $H_{Q1}$ during $t_1$. In that case, it has been shown [40] that the signal is proportional to

$$S(\tau) \propto \frac{2}{3}\sin^2(\theta)\langle\sin^2(2\omega_{HN}\tau)\rangle + \frac{2}{3}\sin^2(2\theta)\langle\sin^4(\omega_{HN}\tau)\rangle \qquad (6)$$

where $\langle...\rangle$ denotes the powder average, and for $^1H$-$\{^{14}N^{DQ}\}$ it is proportional to

$$S(\tau) \propto \frac{2}{3}\sin^4(\theta)\langle\sin^4(\omega_{HN}\tau)\rangle. \qquad (7)$$

For $J$-HMQC experiments, $\omega_{HN} = \pi J_{HN}$, and with HP the maximal transfer efficiency is obtained for $\theta = \pi/2$ and a delay of

$$\tau = 1/(4J_{HN}) \text{ for } ^1H\text{-}\{^{14}N^{SQ}\} \text{ or } 1/(2J_{HN}) \text{ for } ^1H\text{-}\{^{14}N^{DQ}\} \qquad (8)$$

This is contrary to *J*-HMQC experiments with a pair of two spin-1/2 nuclei, where the optimum signal is always observed for $\tau = 1/(2J)$.

## 2.2. $D_1^K$

$^{14}N$ SQ and DQ coherences in $^1H$-$\{^{14}N\}$ HMQC experiments can also be created using rotor-synchronized DANTE schemes, denoted $D_M^K$, which consist of a train of rectangular pulses of length, $\tau_p$, rf-amplitude, $\nu_1$, with identical phases. The length of a $D_M^K$ train is equal to $KT_R$ and $M$ equally spaced pulses are applied during each of these $K$ rotor periods. The $D_M^K$ schemes with $M = 1$ and $M \geq 2$ are referred to as basic and interleaved DANTE sequences, respectively [10-12]. The DANTE excitation profile in the frequency domain consists of a comb of rf-spikelets, all separated from the carrier frequency by a multiple of $M\nu_R$, with amplitude decreasing with the separation from the carrier frequency according to a sinc function. For a $D_1^K$ scheme resonant with the center-band and with individual pulses satisfying Eq.4, the first-order average Hamiltonian can be written as [40,47]

$$\overline{H} = 2\pi\nu_1\tau_p I_x/T_R \qquad (9)$$

This equation is only valid when the evolution under $H_{Q2}$ is negligible during $KT_R$. As the magnitude of $H_{Q2}$ is proportional to $C_Q^2/\nu_{0,14N}$, for large $C_Q$ values this assumption requires high $B_0$ field and MAS frequencies, which limit the losses due to $H_{Q2}$ during $D_1^K$ [40]. When Eq.9 is valid and the excitation and reconversion of $D_1^K$ schemes are rotor-synchronized, *i.e.*, separated by an integer multiple of rotor periods, the signals of $^1H$-$\{^{14}N^{SQ}\}$ and $^1H$-$\{^{14}N^{DQ}\}$ HMQC experiments using DANTE schemes are given by Eqs. 6 and 7, respectively, where $\theta = 2\pi K\nu_1\tau_p$, and the optimum delays by Eq.8. In a first approximation, the $D_1^K$ train can hence be viewed as a HP with an effective rf-field of $K\nu_1$. If $S_{FWHM}$ denotes the offset width of each excitation **S**pikelet, and $E_{FWHM}$ that of the global **E**nvelope excitation, then these values are equal to [40]

$$S_{FWHM} \approx 1.1/KT_R; \quad E_{FWHM} \approx 1.1/\tau_p \qquad (10)$$

In order to obtain an excitation profile that is robust with respect to the offset, $S_{FWHM}$ must be large and hence K must be small. This implies that most of the time, except when the isotropic frequency range, $\Delta\nu_{14N,iso}$, is small, the largest possible rf-field should be used. It is common that several different $^{14}N$ species exist in the same sample, all with different $C_Q$ values and isotropic resonances, $\nu_{14N,iso}$. Therefore, as a starting point for the experimental optimization, the number of rotor periods should accommodate the isotropic frequency range, and the individual pulse length should excite the majority of the static linewidth for the maximum $C_Q$ value

$$K \approx 1/(T_R*\Delta\nu_{14N,iso}) \text{ and } \tau_p \leq 0.6/C_{Q,max} \qquad (11)$$

However, these simple rules are mainly applicable in the case of weak $H_{Q2}$ effects, as discussed later.

It has been shown (i) that $^1H$-$\{^{14}N^{DQ}\}$ HMQC experiments are less efficient than the $^1H$-$\{^{14}N^{SQ}\}$ ones, and (ii) that HMQC with $D_1^K$ is more efficient than with $D_M^K$ (M > 1) [39,40,45]. For these reasons, we only studied here $^1H$-$\{^{14}N^{SQ}\}$ HMQC sequences using $D_1^K$ trains.

## 2.3. SLP

The HP scheme is based on the assumption of a simultaneous excitation of all crystallites, which is why the maximum possible rf-field strength is then used. Recently, another scheme has been proposed, based on the rotor-driven sequential excitation of the crystallites, which occurs when the time-dependent quadrupole interaction becomes small as a result of sample rotation [39]. This principle leads to the use of selective long pulses (SLP). It must be noted that the name of SLP may appear to be misleading because, as we will see below, this pulse is robust towards offsets (Figs.**6** and **12**). This is contrary to the way SLP are used in liquids. However, in MAS NMR it has very recently been demonstrated that, contrary to HP and $D_1^K$ pulses that excite numerous sidebands, SLP only excites the sideband that coincides with the carrier frequency (Figs.**9d** and **14d** in [59]). As a result, SLP can be considered as selective when only observing the sideband combs. The relatively high efficiency of SLP comes from the fact that all crystallites in the rotor provide a signal, which always occurs when their interaction frequency matches that of the carrier irradiation due to sample rotation. This is contrary to HP and $D_1^K$ excitations, where a large part of the sample gives a small signal due to rf-inhomogeneity. As a consequence, the SLP signal is always at the spinning sideband that is the closest to the carrier frequency, not at the isotropic chemical shift, as shown in Fig.**9d** in [59]. The selective excitation of a single spinning sideband of the $^{14}N^{SQ}$ spectrum requires a sufficiently long pulse with low rf-amplitude: $\tau_p > T_R$ and $\nu_1 < \nu_R$. We employed here rectangular pulses satisfying these conditions. The evolution of the density operator between time points separated by several rotor periods can be described by an average Hamiltonian. For a SLP with phase $\phi_p$ and resonant with the center-band of the $^{14}N^{SQ}$ spectrum, its first-order value is given by [41]

$$\bar{H}^{(1)} = 2\pi\nu_1 A_0 R_z(\phi_p) U_Q(\xi_0)^{-1} I_x U_Q(\xi_0) R_z(\phi_p)^{-1} \tag{12}$$

$A_0$ and $\xi_0$ are the intensity and the phase of the center-band of the $^{14}N^{SQ}$ spinning sideband manifold. $A_0$ does not depend on the $\gamma_{PR}$ Euler angle specifying the orientation of the electric field gradient tensor with respect to the rotor-fixed frame, but depends on the two other Euler angles $\{\alpha_{PR}, \beta_{PR}\}$. The scaling of the rf-field strength by $A_0$ stems from the fact that the instantaneous resonance frequency of the nucleus changes rapidly during the pulse and hence the rf-field rotates the magnetization only during a part of the pulse. $R_z(\phi_p)$ is the rotation operator of the $^{14}N$ spin about the z-axis through angle $\phi_p$ and $U_Q(\xi_0)$ is the quadrupolar propagator defined as:

$$U_Q(\xi_0) = \exp\left[-i\xi_0 \frac{Q_z}{3}\right] \tag{13}$$

When considering only the Hamiltonian given by Eq.12 during SLP, it can be shown that the $^1H$-$\{^{14}N^{SQ}\}$ HMQC signal is proportional to

$$S(\tau) \propto \frac{2}{3}\langle \sin^2(\theta)\sin^2(2\omega_{HN}\tau)\rangle + \frac{2}{3}\langle \sin^2(2\theta)\sin^4(\omega_{HN}\tau)\rangle \tag{14}$$

and that of $^1H$-$\{^{14}N^{DQ}\}$ to

$$S(\tau) \propto \frac{2}{3}\langle \sin^4(\theta)\sin^4(\omega_{HN}\tau)\rangle \tag{15}$$

where $\theta = 2\pi A_0 \nu_1 \tau_p$ depends on the Euler angles $\{\alpha_{PR}, \beta_{PR}\}$ through $A_0$.

Furthermore, for a SLP with a phase $\phi_p$ and resonant with the center-band of the $^{14}N^{SQ}$ spectrum, the second-order average Hamiltonian is given by [41]

$$\bar{H}^{(2)} = \frac{\omega_1^2}{4\omega_R} d\left[Q_z + R_z(\phi_p) D_x R_z(\phi_p)^{-1}\right] \tag{16}$$

with

$$d = \sum_{p=-\infty, p\neq 0}^{+\infty} \frac{A_p^2 - A_{-p}^2 - 2A_p A_0 \cos(\xi_p - \xi_0) + 2A_{-p} A_0 \cos(\xi_{-p} - \xi_0)}{p} \tag{17}$$

where $A_p$ and $\xi_p$ denote the intensity and the phase of the $^{14}N^{SQ}$ $p^{th}$-order spinning sideband manifold. In Eq.16, $\bar{H}^{(2)}$ contains $Q_z$ and $D_x$ terms, which commute [46]. The latter can convert the $I_z$ operator in Eq.1 into $D_y$ one, which represents the $^{14}N^{DQ}$ coherences along the y-axis [39] and hence, this term contributes to the signal of $^1H$-$\{^{14}N^{DQ}\}$ HMQC. However, the double-quantum term of $\bar{H}^{(2)}$ in Eq.16 does not commute with $\bar{H}^{(1)}$, which can thus perturb the creation of $^{14}N^{DQ}$ coherences in $^1$H-$\{^{14}N^{DQ}\}$ HMQC experiments, by creating unwanted $^{14}N^{SQ}$ coherences.

## 2.4. XiX

This issue can be circumvented by the use of the XiX scheme, which is constructed as a train of $2N$ SLP pulses of the same length $T_R$, with alternating phase $x$ and $-x$ to eliminate $\bar{H}^{(1)}$ [41]. For XiX scheme resonant with the center-band, $\bar{H}^{(2)}$ is given by Eq.16 with $\phi_P = 0$ and the signal of $^1$H-$\{^{14}N^{DQ}\}$ HMQC is equal to

$$S(\tau) \propto \frac{2}{3}\langle \sin^2(\theta)\sin^2(2\omega_{HN}\tau)\rangle \quad (18)$$

with $\theta = 2\pi N d(\nu_1/\nu_R)^2$ [41]. Hence, for $^1$H-$\{^{14}N^{DQ}\}$ $J$-HMQC experiments using XiX schemes, the maximal transfer efficiency is achieved for

$$\tau = 1/(4J_{HN}) \quad (19)$$

## 3.  1D numerical simulations

### 3.1. Simulation parameters

The spin dynamics during the aforementioned pulse sequences were simulated using SIMPSON software [48]. The powder average was calculated using 3864 $\{\alpha_{MR}, \beta_{MR}, \gamma_{MR}\}$ Euler angles that described the orientation of the molecule in the rotor frame. The 966 $\{\alpha_{MR}, \beta_{MR}\}$ pairs were selected according to the REPULSION algorithm [49], whereas the four $\gamma_{MR}$ angles were regularly stepped from 0 to 360°. The simulations were performed for one isolated $^1$H-$^{14}$N spin pair. The $^{14}$N nucleus was subject to $H_{Q1}$ and $H_{Q2}$ with $\{C_Q \text{ (MHz)}, \eta_Q\} = \{1.18, 0.5\}$ or $\{3.21, 0.32\}$, which result in powder patterns with a full width of 1.77 and 4.82 MHz, respectively. For a powder-averaged resonance, two effects are related to $H_{Q2}$, both proportional to $C_Q^2/\nu_{0,14N}$: a crystallite dependent shift which leads to a second-order broadening, and an overall positive quadrupole induced shift (QIS) of the gravity center

$$\nu_{QIS} = (3 + \eta_Q^2)C_Q^2/(40\nu_{0,14N}) \quad (20)$$

At $B_0 = 18.8$ T, $\nu_{QIS} = 1957$ and 13827 Hz, for the two previous cases with $C_Q = 1.18$ and 3.21 MHz, respectively. The former parameters correspond to those of $^{14}$N nucleus of α-glycine [15,50] and are typical of those measured in amino acids. The latter correspond to those of N-acetyl-L-valine (NAV) [51]. Except in Fig.**S4**, no CSA was introduced because the magnitude of this interaction is often negligible with respect to the quadrupole one [52,53] In order to decrease the computing time, the simulations were done with the $J$-HMQC sequence shown in Fig.**1a**, instead of the targeted $D$-HMQC one (Fig.**1b**). This, however, does not change the analysis of the efficiency of the $^{14}$N excitation and reconversion, which is the main topic of this article. Therefore, in the simulations performed at $B_0 = 18.8$ T with $\nu_R = 62.5$ kHz, $^1$H and $^{14}$N nuclei were only scalar coupled with $J_{1H-14N} = 400$ Hz.

We simulated the 1D signals of $^1$H-$\{^{14}N^{SQ}\}$ $J$-HMQC sequences using HP, SLP, $D_1^K$ and XiX schemes as well as $^1$H-$\{^{14}N^{DQ}\}$ $J$-HMQC sequences using SLP and XiX schemes. The rf-field for the $\pi/2$ and $\pi$-pulses on the $^1$H channel was fixed to $\nu_{1,1H} = 100$ kHz and was always applied on-resonance. Except in Fig.**6**, the pulses on $^{14}$N channel were applied on resonance, *i.e.*, their carrier frequency was equal to the sum of the isotropic chemical shift and the QIS. The maximum rf-field on $^{14}$N channel did not exceed 200 kHz in order to be consistent with the rf-power specifications of common MAS probes. For $^1$H-$\{^{14}N^{SQ}\}$ $J$-HMQC simulations, the delays were set to their shortest optimal value of $\tau_J = 1/(4J_{HN}) = 2.5$ ms (Eq.8). For $^1$H-$\{^{14}N^{DQ}\}$ simulations, we employed $\tau_J = 1/(4J_{HN})$ for XiX (Eq.19) and $\tau_J = 1/(2J_{HN})$ for SLP (Eq.8). For these 1D simulations, the $t_1$ delay was fixed to an integer multiple of $T_R$, in order to refocus the evolution under $H_{Q1}$. For HP and SLP, $t_1 = 2T_R$, which is the largest $\tau_p$ value used in the simulations. For the same reason, $t_1$ was fixed to $KT_R$ and $2NT_R$ for $D_1^K$ and XiX schemes, respectively. In all $J$-HMQC simulations, the transfer efficiency was normalized with respect to the $^1H$ signal observed with an ideal spin-echo experiment with Dirac pulses. The maximum efficiency of $^1H$-$\{^{14}N^{SQ}\}$ $J$-HMQC using two ideal Dirac $^{14}N$ $\pi/2$-pulses for one $^1H$-$^{14}N$ spin pair with $C_Q = 0$ is equal to 0.35. This ideal value will be used as a benchmark in the following discussion.

### 3.2. Optimal rf-field amplitude and duration

The simulation-based analysis started by calculating a series of two-dimensional maps showing the transfer efficiency *versus* the rf-field and the duration of excitation and reconversion schemes applied to $^{14}N$ channel on-resonance. In the case of HP or SLP these maps are displayed as ($\nu_1$ *vs.* $\tau_p$), whereas for XiX, they are displayed as ($\nu_1$ *vs.* N). In the case of $D_1^K$, each ($\nu_1$, $\tau_p$) map was calculated for one particular $K$ value. The maps corresponding to the best efficiencies are shown either in the main text (Figs.**2-5**) or in the supplementary information (Figs.**S1** and **S2**).

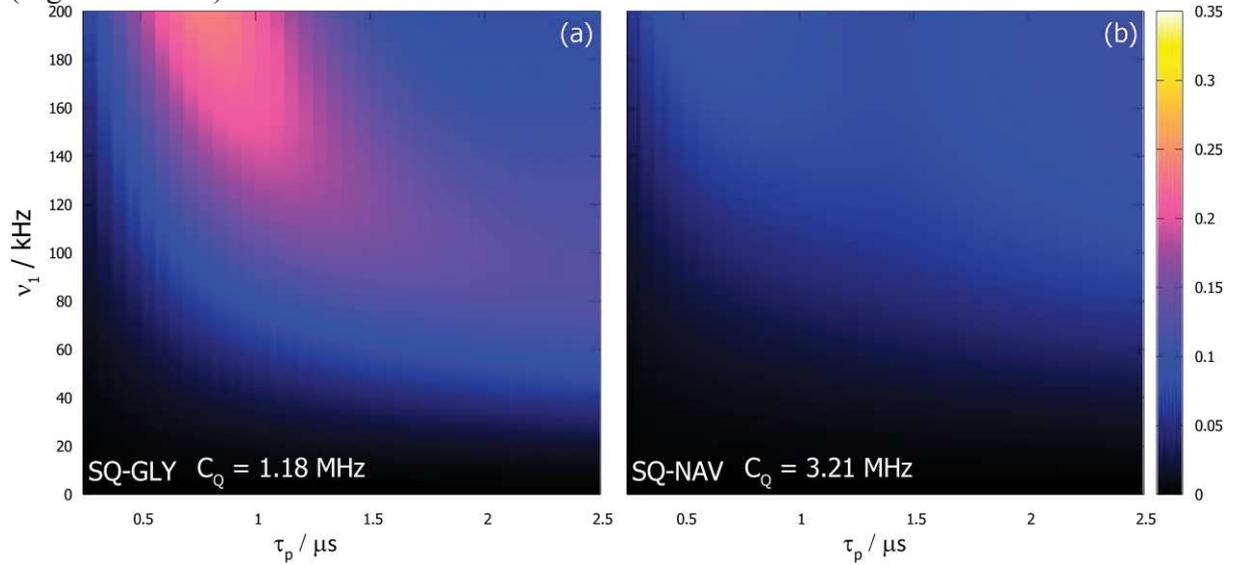

**Fig.2.** Simulated on-resonance efficiency of $^1H$-$\{^{14}N^{SQ}\}$ $J$-HMQC sequence using HP *versus* $\tau_p$ and $\nu_1$ for spin systems corresponding to (a) glycine and (b) NAV.

As seen in Fig.**2a**, the maximal efficiency for $^1H$-$\{^{14}N^{SQ}\}$ $J$-HMQC using HP is achieved for $\tau_p = 0.75$ μs and $\nu_1 = 200$ kHz, which corresponds to $\theta = 54°$. This angle is lower than 90°. Furthermore, $\tau_p = 0.75$ μs does not satisfy Eq.4 for glycine and hence, HP does not produce a uniform excitation of the $^{14}N^{SQ}$ powder pattern. Therefore, the efficiency of the sequence only reaches 0.2, which is 40% smaller than that of $^1H$-$\{^{14}N^{SQ}\}$ $J$-HMQC sequence using Dirac pulses for a $^{14}N$ nucleus not subject to quadrupole interaction. As seen in Fig.**2b**, the efficiency is further decreased for larger $C_Q$.

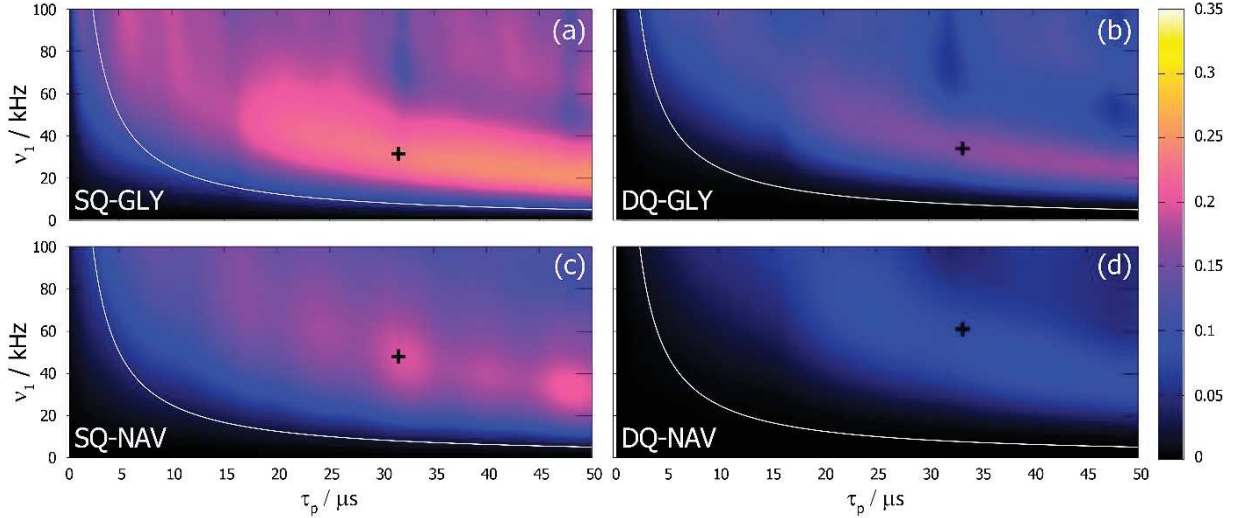

**Fig.3.** Simulated on-resonance efficiency of (a,c) $^1\text{H-}\{^{14}\text{N}^{\text{SQ}}\}$ and (b,d) $^1\text{H-}\{^{14}\text{N}^{\text{DQ}}\}$ $J$-HMQC sequence using SLP *versus* $\tau_p$ and $\nu_1$ for spin systems corresponding to (a,b) glycine and (c,d) NAV. The $\tau_p$ and $\nu_1$ values corresponding to $\pi/2$-pulses are plotted as solid white lines.

For $^1\text{H-}\{^{14}\text{N}^{\text{SQ}}\}$ $J$-HMQC using SLP (Fig.**3a** and **c**), the maximal transfer efficiency is obtained for $\tau_p \approx 2T_R$ and it reaches 0.23 with $\nu_1 \approx 35$ kHz for glycine and 0.20 with $\nu_1 \approx 50$ kHz for NAV. Those efficiencies are *ca.* the same and 100% higher than those obtained for glycine and NAV, respectively, using HP with $\nu_1 = 200$ kHz. These results indicate that (i) SLP is more efficient than HP for the excitation and the reconversion of $^{14}\text{N}^{\text{SQ}}$ coherences, notably for $^{14}\text{N}$ nuclei subject to large quadrupole interaction, and (ii) SLP requires much lower rf-field than HP. For $^1\text{H-}\{^{14}\text{N}^{\text{DQ}}\}$ $J$-HMQC using SLP, maximal transfer efficiencies are obtained under similar conditions than for $^1\text{H-}\{^{14}\text{N}^{\text{SQ}}\}$ (Fig.**3b** and **d**). However, the excitation and the reconversion of $^{14}\text{N}^{\text{DQ}}$ coherences are significantly less efficient than those of $^{14}\text{N}^{\text{SQ}}$ ones.

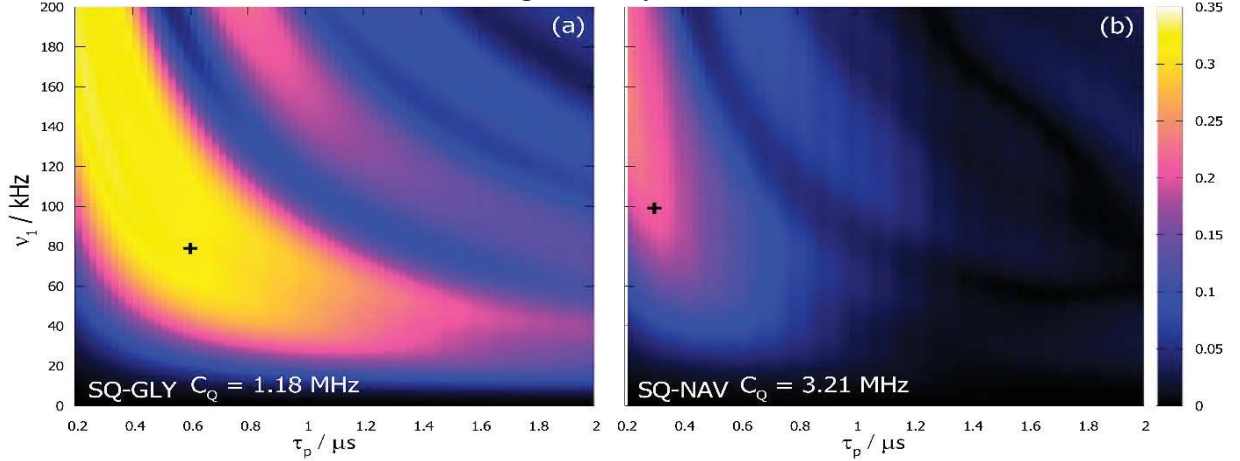

**Fig.4.** Simulated on-resonance efficiency of $^1\text{H-}\{^{14}\text{N}^{\text{SQ}}\}$ $J$-HMQC sequence using $D_1^K$ *versus* $\tau_p$ and $\nu_1$ for spin systems corresponding to (a) glycine with $K = 5$ and (b) NAV with $K = 6$.

For $^1\text{H-}\{^{14}\text{N}^{\text{SQ}}\}$ $J$-HMQC using $D_1^5$ applied to glycine, a maximal efficiency of 0.32 is obtained for rf-fields down to $\nu_1 = 80$ kHz. For such rf-field, the optimal $\tau_p$ length is 0.6 μs, which corresponds to $\theta \approx 86°$. This angle is close to the theoretical optimal angle of 90°, whereas the achieved efficiency reaches almost the theoretical one of 0.35. This high efficiency indicates small losses due to $H_{Q2}$. For NAV, $^1\text{H-}\{^{14}\text{N}^{\text{SQ}}\}$ $J$-HMQC with $D_1^6$ yields a maximal efficiency of 0.21 for $\tau_p \approx 0.3$ μs and $\nu_1 = 100$ kHz, *i.e.*, $\theta \approx 65°$. The reduced efficiency and tilt angle stem from faster $H_{Q2}$ dephasings for NAV than glycine since $H_{Q2}$ is proportional to $C_Q^2$. As a simple rule, when $H_{Q2}$ dephasings are small, $T_R\nu_{QIS} \leq 0.05$ (e.g. 0.032 for glycine), the optimal total pulse duration corresponds to θ ≈ 90°, but it decreases when it is not the case (e.g. 0.22 for NAV).

The simulations showing the simulated efficiency of $^1$H-{$^{14}$N$^{SQ}$} $J$-HMQC with D$_1^K$ scheme and $3 \leq K \leq 8$ for a spin system corresponding to glycine or NAV are shown in the Supplementary Information (Figs.**S1** or **S2**, respectively). Practically, the D$_1^K$ optimization is performed by fixing $\nu_1$ at its maximum value (80 kHz with our MAS probe), and by optimizing $\tau_p$ for various $K$ values.

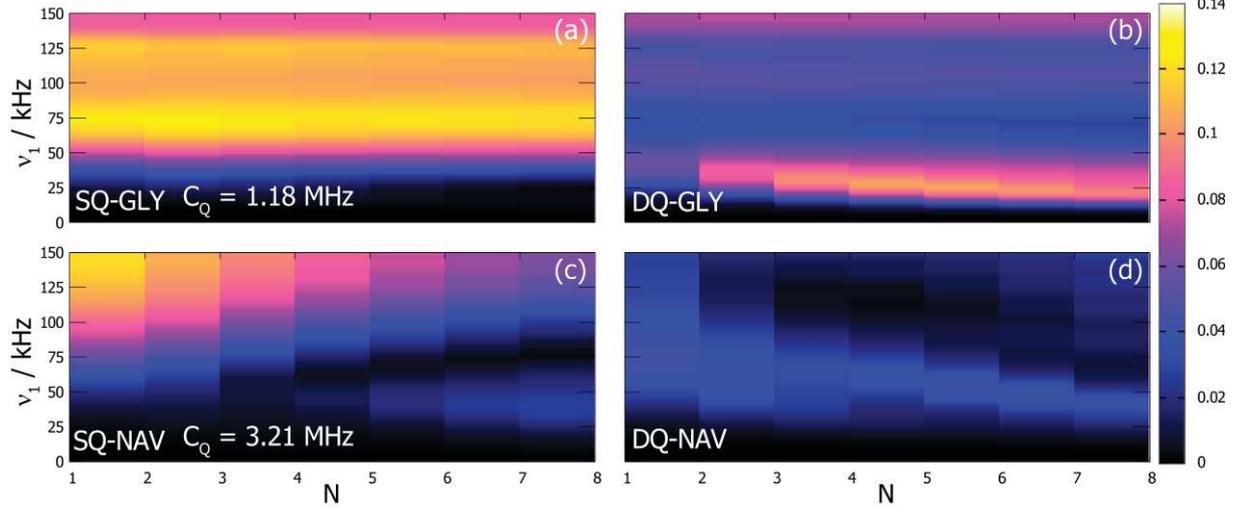

**Fig.5.** Simulated on-resonance efficiency of (a,c) $^1$H-{$^{14}$N$^{SQ}$} and (b,d) $^1$H-{$^{14}$N$^{DQ}$} $J$-HMQC sequence using XiX *versus* $N$ and $\nu_1$ for spin systems corresponding to (a,b) glycine and (c,d) NAV.

Fig.**5b** shows that maximal efficiency for $^1$H-{$^{14}$N$^{DQ}$} $J$-HMQC with XiX is obtained for $\nu_1$ = 30 kHz and $N = 4$ for glycine. For NAV, the maximal efficiency is obtained for smaller $N$ values because of losses due to H$_{Q2}$ (Fig.**5d**). Furthermore, for both glycine and NAV, XiX results in 45% lower efficiency for $^1$H-{$^{14}$N$^{DQ}$} $J$-HMQC than with SLP. Fig.**5a** and **c** indicate that XiX can also be used to excite and reconvert the $^{14}$N$^{SQ}$ coherences. Shorter $N$ values are required for NAV in order to limit the losses produced by H$_{Q2}$. $^1$H-{$^{14}$N$^{SQ}$} $J$-HMQC using XiX are twice less efficient than those using SLP for glycine and NAV.

The optimum parameters obtained from the aforementioned maps, as well as their resulting normalized efficiencies, Eff, are summarized in Table 1. These parameters were used in subsequent simulations. No further discussion of the XiX results will be presented in the remainder of this article, because their simulated and experimental efficiency and robustness (not shown) were consistently found to be much smaller than those of the SLP and D$_1^K$ schemes.

Table **1.** Optimum on-resonance simulation parameters used for glycine or NAV.

| Excitation | $\tau_J$ | $\tau_p$ / μs | $\nu_1$ / kHz | Eff |
|---|---|---|---|---|
| *Glycine* | | | | |
| HP SQ | 1/4J | 0.75 | 200 | 0.24 |
| SLP SQ | 1/4J | 32 (2T$_R$) | 30 | 0.23 |
| SLP DQ | 1/2J | 32 (2T$_R$) | 35 | 0.16 |
| XiX SQ | 1/4J | 64 (N = 2) | 75 | 0.13 |
| XiX DQ | 1/4J | 128 (N = 4) | 30 | 0.11 |
| D$_1^K$ SQ | 1/4J | 0.5 (K = 5) | 80 | 0.32 |
| *NAV* | | | | |
| HP SQ | 1/4J | 0.75 | 200 | 0.11 |
| SLP SQ | 1/4J | 32 (2T$_R$) | 50 | 0.20 |
| SLP DQ | 1/2J | 32 (2T$_R$) | 55 | 0.10 |

| XiX SQ | 1/4J | 32 (N = 1) | 140 | 0.13 |
| XiX DQ | 1/4J | 64 (N = 2) | 65 | 0.05 |
| $D_1^K$ SQ | 1/4J | 0.3 (K = 6) | 100 | 0.21 |

### 3.3. Robustness

### 3.3.1. Offset

In addition to the on-resonance sensitivity, another important criterion is the robustness to offset, which is shown in Fig.**6a** for glycine and Fig.**6b** for NAV. The two most robust excitations are those related to SLP. Their widths are the same with SQ or DQ selection and they depend very little upon the $C_Q$ value ($S_{FWHM} \approx 40$ kHz). The $D_1^K$ scheme is less robust ($S_{FWHM} \approx 20$ kHz), and this value is the same for glycine and NAV. These results indicate that the robustness to offset mainly depends on the length of the excitation and reconversion blocks (see $S_{FWHM}$ in Eq.10). The shorter length of SLP with respect to $D_1^K$ explains the higher robustness to offset of the former.

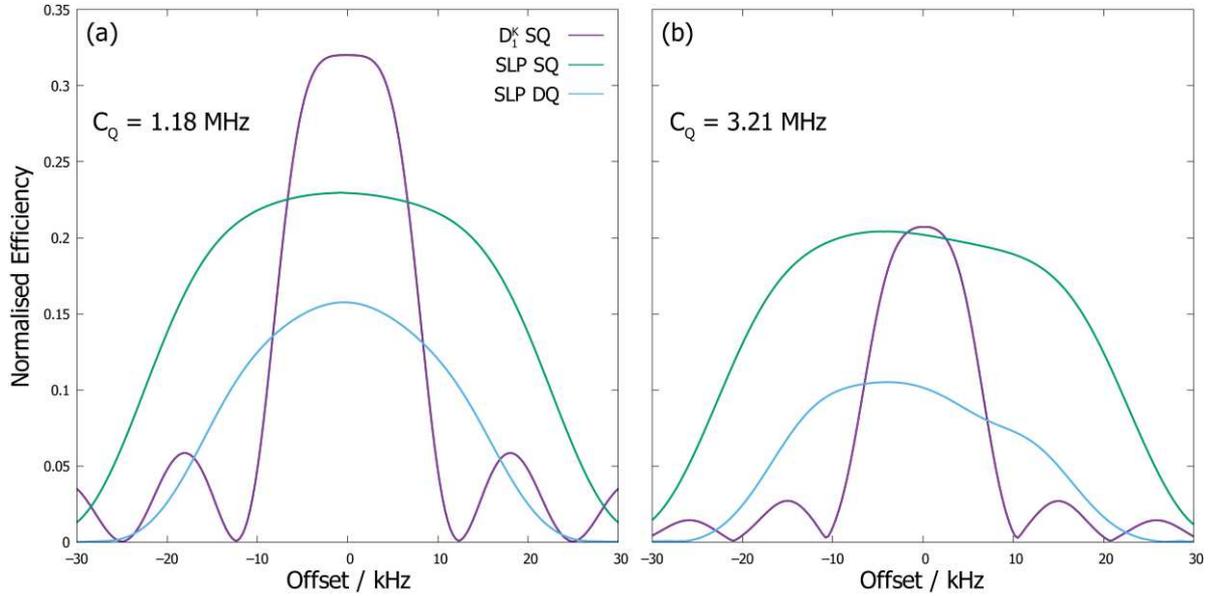

**Fig.6.** Simulated efficiency of $^1$H-{$^{14}$N$^{SQ}$} and $^1$H-{$^{14}$N$^{DQ}$} $J$-HMQC sequences using SLP and $^1$H-{$^{14}$N$^{SQ}$} with $D_1^K$ with respect to $^{14}$N offset for spin systems corresponding to (a) glycine and (b) NAV. The rf parameters ($\nu_1$, $\tau_p$) are indicated with + on Figs.**2-4**.

### 3.3.2. Rf-inhomogeneity

Previous simulations were performed with one particular rf-field. However, there exists an inherent distribution of rf-fields in the coil of NMR probes, mostly along the axis of the solenoid coil in MAS probes. Typically, this amplitude is maximal at the center of the coil and it drops by *ca*. 50 % near the edges [54–57].

For SLP, the simulations displayed in Fig.**7** show that, independently of the $C_Q$ value, (i) the optimal rf-field is equal to *ca*. 50-80 % of $\nu_R$, and (ii) the relative widths of these curves, *i.e.,* the FWHM divided by the optimum rf-field, which are the relevant values with respect to rf-inhomogeneity, are approximately the same with SQ or DQ selection.

With $D_1^K$ trains, the rf-requirement is higher than with SLP, but the efficiency is larger with small $C_Q$ values. Conversely for large $C_Q$ values, the losses due to $H_{Q2}$ decrease the efficiency.

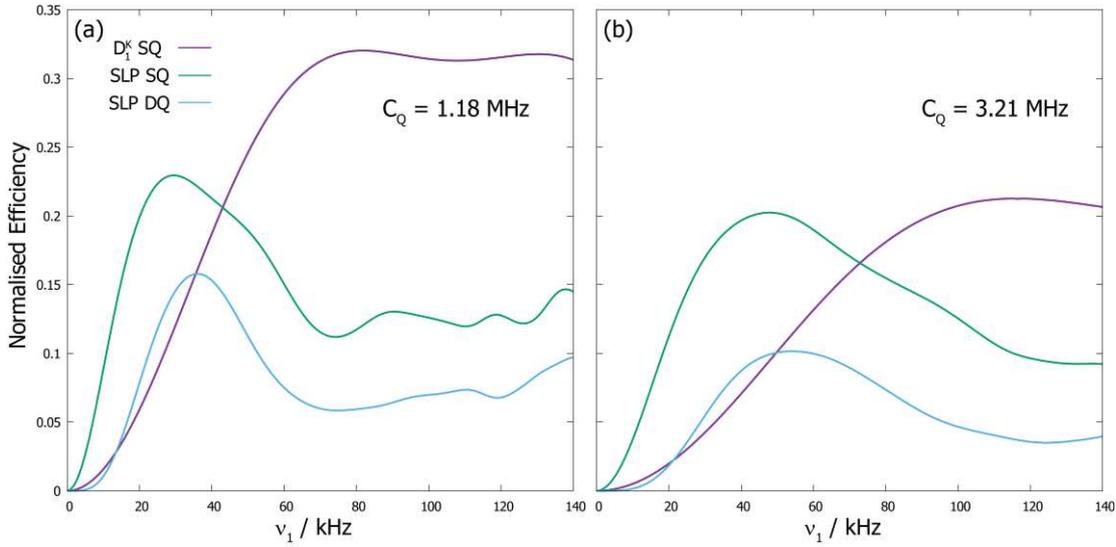

**Fig.7.** Simulated on-resonance efficiency of $^1$H-$\{^{14}$N$^{SQ}\}$ and $^1$H-$\{^{14}$N$^{DQ}\}$ *J*-HMQC sequences using SLP and $^1$H-$\{^{14}$N$^{SQ}\}$ *J*-HMQC sequences using $D_1^K$ with respect to respect to $^{14}$N rf-field strength for spin systems corresponding to (a) glycine and (b) NAV. The rf-parameters ($\nu_1$, $\tau_p$) are indicated with + on Figs.**2-4**, with $D_1^5$ in (a) and $D_1^6$ in (b).

### 3.3.3. Spinning speed and CSA

The robustness with respect to spinning speed and CSA (Figs.**S3** and **S4**, respectively) was also tested. The simulations show that SLP depends little upon these two parameters, but that $D_1^K$ is sensitive with respect to spinning speed fluctuations. However, modern MAS speed controllers can provide sufficiently stable MAS frequency to allow the use of $D_1^K$ schemes.

### 3.3.4. Quadrupole interaction

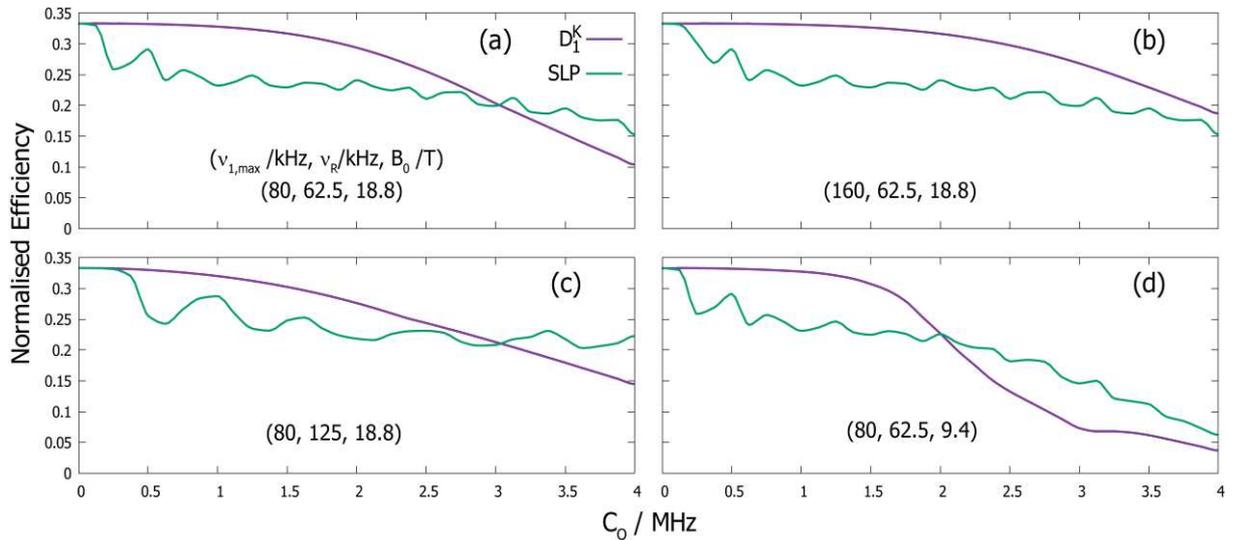

**Fig.8.** Simulated on-resonance optimized efficiency of $^1$H-$\{^{14}$N$^{SQ}\}$ *J*-HMQC sequences using SLP and $D_1^K$ (green and purple curves, respectively) for a series of quadrupole parameters corresponding to $\eta_Q = 0.5$ and $C_Q = 0$-4 MHz, incremented in steps of 125 kHz for ($\nu_{1,max}$/kHz, $\nu_R$/kHz, $B_0$/T) = (a) (80, 62.5, 18.8); (b) (160, 62.5, 18.8); (c) (80, 125, 18.8) and (d) (80, 62.5, 9.4). For SLP, the $\tau_p$ range investigated was 1.5-2.5$T_R$. For $D_1^K$, the investigated range of the total pulse duration, $K\tau_p$, is 0-10 $\mu$s, with $K$ ranging from 3 to 8, as also shown in Fig.**S1** and **S2**.

The last aspect analyzed *via* simulations was the robustness with respect to the $C_Q$ value. This is an important criterion as this parameter can vary between 0 and 7 MHz, depending on the local environment of $^{14}$N nucleus [5,7]. To that end, the optimum efficiency of the two most efficient methods, SLP and $D_1^K$ with SQ selection, was calculated for a series of quadrupole parameters corresponding to $C_Q$ = 0-4 MHz. These simulations were performed with the following sets of parameters: $(\nu_{1,max}/kHz, \nu_R/kHz, B_0/T)$ = (80, 62.5, 18.8); (160, 62.5, 18.8); (80, 125, 18.8) and (80, 62.5, 9.4), and the corresponding results are shown in Fig.**8a-d**, respectively. These results show that $D_1^K$ on-resonance excitation is more efficient than SLP for small and moderate $C_Q$ values and that it is the contrary for large $C_Q$ values. It can be seen that the efficiency of $D_1^K$ increases: (i) with $\nu_1$ at moderate to high $C_Q$ value (compare Figs.**8a** and **b**), and (ii) with $\nu_R$ and $B_0$ at high $C_Q$ value (compare Figs.**8a, c** and **d**). Indeed, a large rf field allows the use of shorter $D_1^K$ trains, *i.e.*, smaller $K$ values (compare Figs.**9a** and **b**), which limits the losses due to $H_{Q2}$, which are proportional to $C_Q^2$. Similarly high MAS frequency decreases the length, $KT_R$, of $D_1^K$ train, hence reducing losses due to $H_{Q2}$. Finally, as $H_{Q2}$ is inversely proportional to $B_0$, losses produced by this interaction are smaller at high field. By contrast, the on-resonance SLP excitation exhibits very little dependence upon these parameters, and hence the range of $C_Q$ values where $D_1^K$ is more efficient than SLP extends with increasing $\nu_1$, $\nu_R$ and $B_0$ values. However, it must be pointed out that the robustness to offset of SLP is much larger than that of $D_1^K$ (Fig.**6**), and that this robustness further increases with faster spinning speeds (Fig.**5** of [39]). Therefore, for small or moderate $C_Q$ values, the choice in between these two schemes also depends on the range of isotropic resonance frequencies.

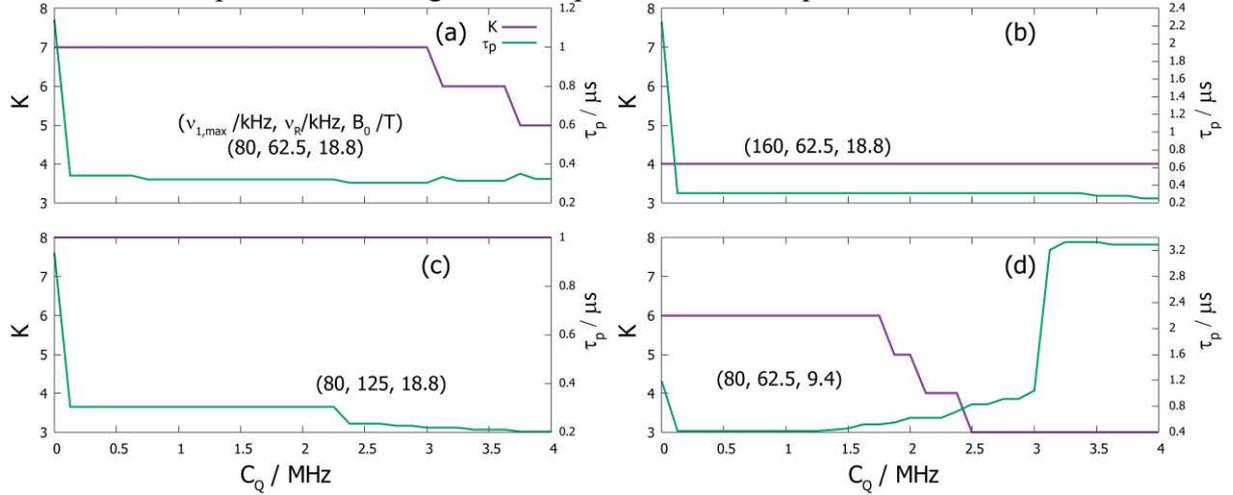

**Fig.9.** $\tau_p$ and $K$ values, green and purple curves respectively, yielding optimal simulated efficiency for $^1$H-{$^{14}$N$^{SQ}$} $J$-HMQC with $D_1^K$ for a series of quadrupole parameters corresponding to $\eta_Q$ = 0.5 and $C_Q$ = 0-4 MHz, incremented in steps of 125 kHz for $(\nu_{1,max}/kHz, \nu_R/kHz, B_0/T)$ = (a) (80, 62.5, 18.8); (b) (160, 62.5, 18.8); (c) (80, 125, 18.8) and (d) (80, 62.5, 9.4).

Fig.**9a-d** plots the variations in optimal $\tau_p$ and $K$ values of the $D_1^K$ scheme with respect to $C_Q$, over the same range of parameters as for Fig.**8**. These results show that, in general, to obtain a maximum efficiency with $D_1^K$, the higher the $C_Q$ value, the shorter the train that should be used. Such an arrangement is effective because it limits the signal decay due to $H_{Q2}$ dephasing [40]. This leads to a decrease of $K$ with large $C_Q$ values in Fig.**9a** and **d**. It can be observed that, at lower $B_0$ field (Fig.**9d**), the lengths of $\tau_p$ at high $C_Q$ are noticeably longer than the equivalent values at 18.8 T. Long pulses allow for reduction in the length of $D_2^K$ train, *i.e.*, the $K$ value, hence limiting the losses due to $H_{Q2}$. Regardless of which values are used, excitation of high $C_Q$ values at low magnetic field is likely to be inefficient with $D_1^K$.

It should be noted that the data presented in Figs.**8** and **9** were sorted computationally and, as such, represent theoretical, numerical maxima. These efficiencies often only differ after several decimals and so the choice between, for example, one $K$ value and another, may vary from a practical perspective.

## 4. Experimental results for $^1$H-{$^{14}$N} $D$-HMQC

### 4.1. Experimental conditions

All experiments were performed at $B_0$ = 18.8 T on a Bruker standard bore superconducting magnet equipped with an AV-IV console. Two model samples were selected for this study: L-histidine HCl and N-acetyl-L-valine, referred to His and NAV, and obtained from Sigma-Aldrich and Alfa Aesar, respectively, and used as-received. Their $^1$H-{$^{14}$N$^{SQ}$} and $^1$H-{$^{14}$N$^{DQ}$} spectra were acquired with the $D$-HMQC sequence depicted in Fig.**1b**, which was employed since it is more sensitive than the $J$-HMQC variant for solids. The samples were packed into 1.3 mm rotors and spun under MAS at $v_R$ = 62.5 kHz, using a standard HX MAS probe. SR4$_1^2$ recoupling with $v_1$ = 2$v_R$ = 125 kHz was applied on $^1$H channel during the $\tau_D$ delays to reintroduce the $^{14}$N-$^1$H dipolar couplings. The $^1$H rf-field for the $\pi$/2 and $\pi$-pulses was set to $v_1$ $\approx$ 116 kHz. The rf-field on the $^{14}$N channel delivered by this probe is limited to $v_1$ $\approx$ 80 kHz. HP, D$_1^K$ trains and SLP were employed to excite and reconvert the $^{14}$N SQ or DQ coherences. The position of $^{14}$N carrier frequency for His is indicated in Fig.**10**. For NAV, the $^{14}$N carrier frequency was resonant with the NH peak. In Fig.**12**, the $^{14}$N carrier frequency was varied. For 1D experiment, the $t_1$ delay was fixed to its minimum rotor-synchronized value equal either to 2$T_R$ for rectangular pulses or $KT_R$ for D$_1^K$ trains. All 2D spectra were acquired using $t_1$ period equal to an integer multiple of $T_R$. This rotor-synchronization allows for refocusing of the broadening by H$_{Q1}$ for $^1$H-{$^{14}$N$^{SQ}$} $D$-HMQC spectra. The spectra were acquired with a relaxation delay, $\tau_{RD}$ = 1 s. The other experimental parameters are given in Table 2 as well as in the figure captions. Chemical shift referencing was carried out using solid samples of NH$_4$Cl ($\delta_{14N}$ = 0 ppm) and His ($\delta_{1H}$ = 8.2 ppm for N$^\tau$H). His has three $^{14}$N species that possess the following NMR parameters: $\delta_{iso}$ = 4.0/146.6/133.4 ppm, $C_Q$ = 1.26/1.46/1.29 MHz and $\eta_Q$ = 0.36/0.30/0.97, for NH$_3$, N$^\pi$H and N$^\tau$H, respectively [**58**]. When taking into account their $v_{QIS}$ values, these peaks appear at $-17200$, $-8500$ and $-9500$ Hz, respectively. NAV has only one species with NMR parameters: $\delta_{iso}$ = 127.7 ppm, $C_Q$ = 3.21 MHz and $\eta_Q$ = 0.32 [**51**]. Its $^{14}$N resonance appears at $-2000$ Hz on the spectra presented in this work.

### 4.2. Optimized parameters

In this part, we present the results observed for His and NAV that were recorded either with conventional HP (SQ, $v_1$ = 80 kHz), D$_1^K$ (SQ, $v_1$ = 40 or 80 kHz) or SLP (SQ and DQ) with optimized $v_1$ values. In Table **2**, it can be observed that the optimized SLP length, $\tau_p$ = 20-28 $\mu$s $\approx$ 1.5$T_R$, depends very little upon the sample and hence on the $C_Q$ value, and that these values are slightly smaller than the simulated one, $\tau_p$ $\approx$ 32 $\mu$s = 2$T_R$. This small decrease is attributed to experimental losses, such as those due to $^1$H-$^1$H dipolar couplings that were not included in the simulations. Concerning the D$_1^K$ scheme, the pulse length with $v_1$ = 80 kHz is longer than in the simulations: 1 instead of 0.6 $\mu$s for His and 0.5 instead of 0.3 $\mu$s for NAV, whereas the optimal $K$ values are smaller: 4 instead of 5 and 6 for His and NAV, respectively. These smaller $K$ values stem from experimental losses, such as those due to $^1$H-$^1$H dipolar couplings, which were not included in the simulations.

Table **2**. Optimal experimental parameters used for His or NAV.

| Excitation | $\tau_D$ / µs | $\tau_p$ / µs | $\nu_1$ / kHz |
|---|---|---|---|
| *His* | | | |
| HP    SQ | 140 | 1 | 80 |
| SLP SQ | 140 | 24 | 40 |
| SLP DQ | 160 | 20 | 53 |
| $D_1^K$   SQ | 140 | 1 ($K = 6$) | 40 |
| $D_1^K$   SQ | 140 | 1 ($K = 4$) | 80 |
| *NAV* | | | |
| HP    SQ | | 1 | 80 |
| SLP SQ | 170 | 28 | 45 |
| SLP DQ | 130 | 28 | 64 |
| $D_1^K$   SQ | 140 | 2  ($K = 3$) | 40 |
| $D_1^K$   SQ | 140 | 0.5 ($K = 4$) | 80 |

## 4.3. Efficiencies

$^1$H-$\{^{14}$N$\}$ *D*-HMQC spectra were acquired with the aforementioned optimized parameters from Table **2**.

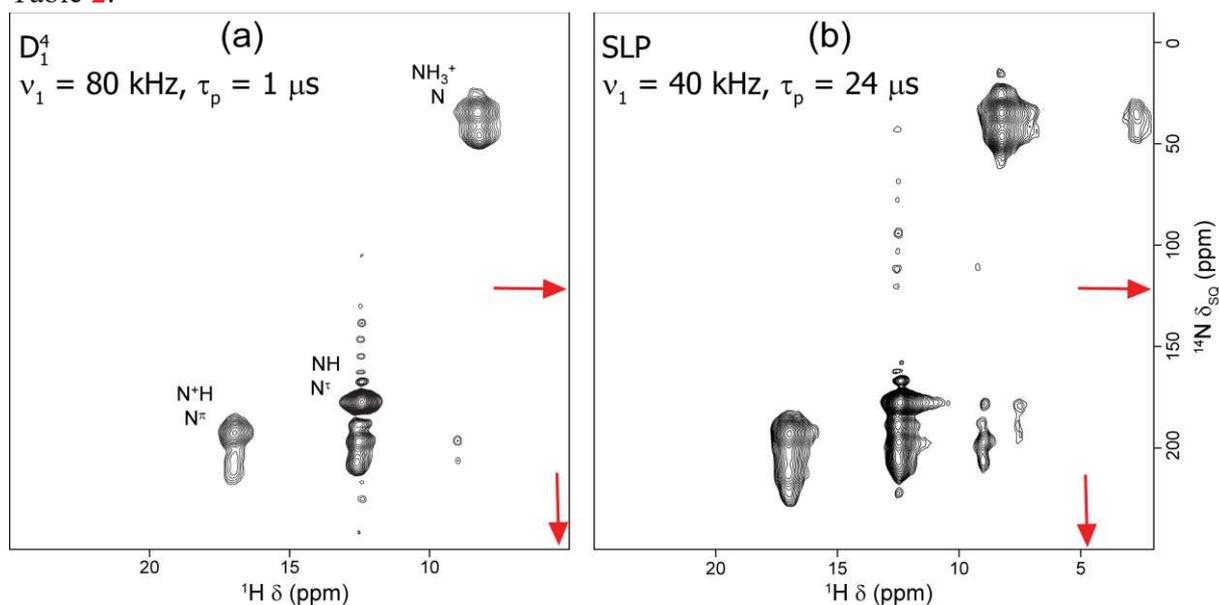

**Fig.10.** $^1$H-$\{^{14}$N$^{SQ}\}$ *D*-HMQC spectra of His acquired using (a) $D_1^4$ train and (b) SLP excitation, with $B_0 = 18.8$ T, $\nu_1 = 80$ ($D_1^4$) or 40 kHz (SLP), $\nu_R = 62.5$ kHz. The 2D spectra result from averaging 8 transients for each of the 250 $t_1$ increments, resulting in a total experiment time of 33 min. Red arrows indicate the positions of the carrier frequencies.

Fig.**10a** and **b** demonstrate that such 2D spectra can be recorded with a good efficiency for His with experimentally achievable rf-fields of $\nu_1 = 80$ or 40 kHz, with SQ selection and either $D_1^K$ or SLP excitation, respectively.

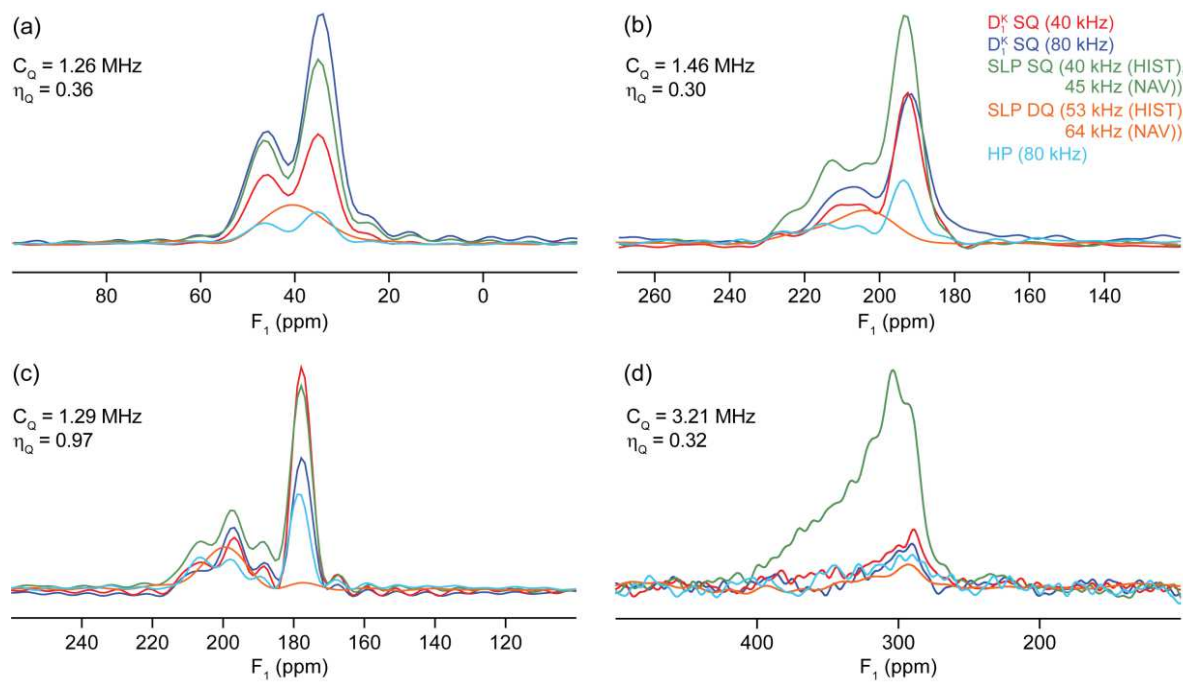

**Fig.11.** $F_1$ slices from 2D $^1$H-$\{^{14}$N$\}$ *D*-HMQC spectra for (a) NH$_3$, (b) N$^\pi$H, (c) N$^\tau$H of His and (d) NH of NAV, with $B_0$ = 18.8 T, $\nu_R$ = 62.5 kHz, and $\nu_1$ as indicated in (b). In all cases, SQ line-shapes have been scaled up by 2 with respect to DQ ones. Additionally, the DQ ppm scale has been divided by two to allow for comparison with SQ slices. The 2D spectra result from averaging 8 (SQ) or 16 (DQ) transients for each of 250 (SQ) or 124 (DQ) $t_1$ increments, resulting in a total experiment time of 33 min.

The $F_1$ slices of the 2D spectra for NH$_3$, N$^\pi$H, N$^\tau$H of His and NAV, are shown in Figs.**11a**, **b**, **c** and **d**, respectively. It can be seen that the S/N ratio observed with NAV is smaller than with His. The $^{14}$N MAS linewidth of NAV is *ca.* 5-6 times larger than those of His. However, this larger indirect linewidth should not significantly affect the sensitivity because it allows for the use of fewer t$_1$ increments. Therefore, the main cause of the reduced HMQC efficiency in NAV is likely to be brought about by the increase in $C_Q$ value (Fig.**8**), and possibly a reduction in the $^1$H-$^{14}$N dipolar coupling or $^1$H coherence lifetime.

It is evident that, even with the highest available rf-field, HP excitation is inefficient, especially for large $C_Q$ values, as it is the case for NAV. With respect to SQ selection, DQ always leads to a large decrease in efficiency, as previously shown *via* simulations. Moreover, no resolution enhancement is observed with DQ selection over the SQ one, indicating that in the latter case the spinning speed was stable and the magic angle was well-adjusted, and hence, the broadening of SQ coherences by H$_{Q1}$ was properly refocused. D$_1^K$ excitation can provide almost the same level of efficiency with $\nu_1$ = 40 or 80 kHz, because the increased value of K$\tau_p$ is able to compensate for the decrease in rf-field strength, a finding which is in agreement with previous studies [12,40]. As seen in Fig.**11a-c**, the relative efficiency of SLP and D$_1^K$ for His depends on the observed site. However, D$_1^K$ is not efficient for large C$_Q$ values, as shown in Fig.**11d** for NAV, owing to large losses due to H$_{Q2}$.

## 4.4. Robustness to offset of SLP

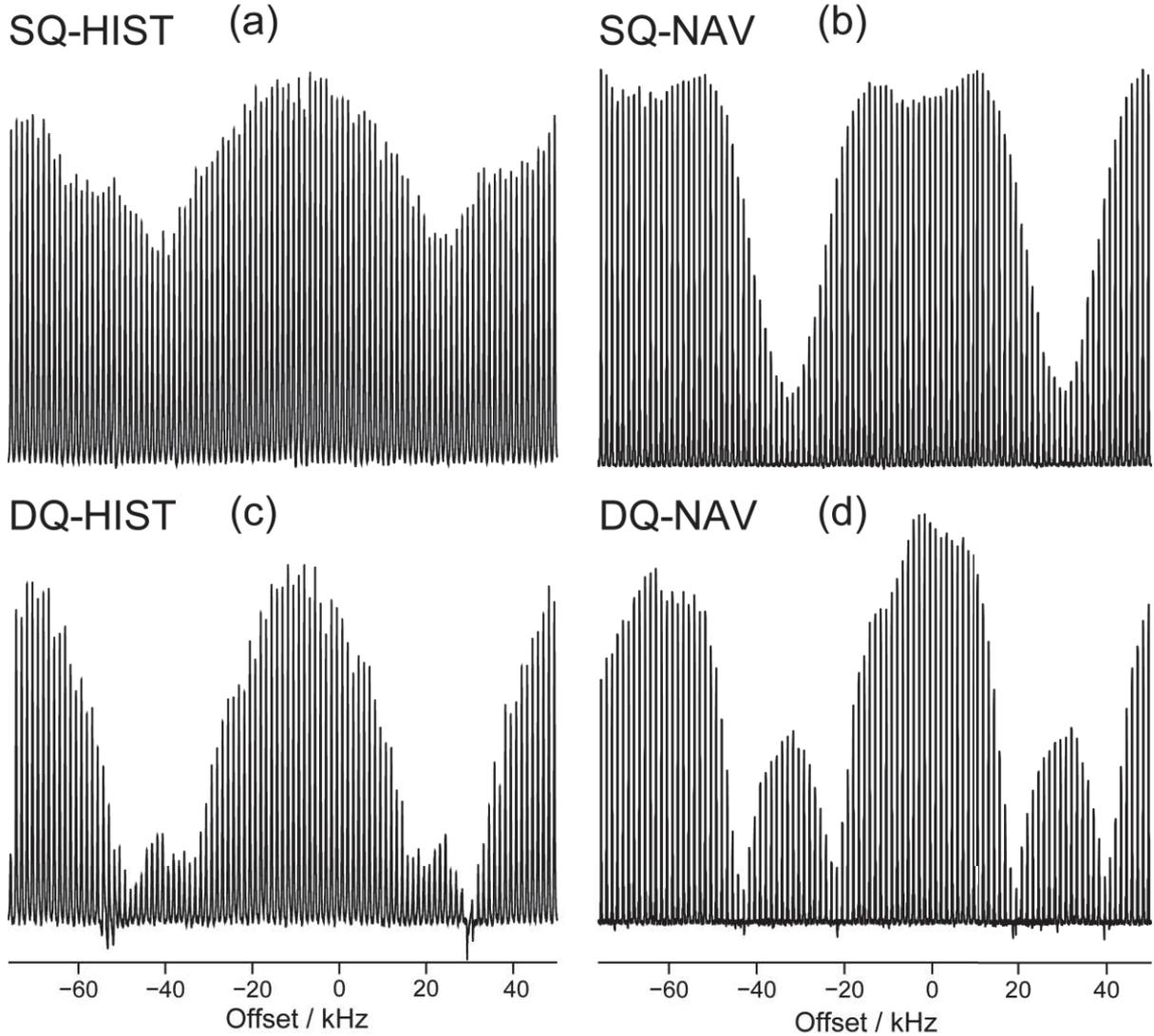

**Fig.12.** Experimental (a,b) $^1$H-{$^{14}$N$^{SQ}$} and (c,d) $^1$H-{$^{14}$N$^{DQ}$} *D*-HMQC 1D signal *versus* $^{14}$N offset with SLP of (a,c) N$^{\varepsilon}$H of His and (b,d) NH of NAV, at $B_0 = 18.8$ T with $\nu_R = 62.5$ kHz using $\nu_1 = 40$ (a,c) or (b,d) 45 kHz.

Fig.**12** shows the sensitivity to $^{14}$N offset observed with SLP and SQ or DQ selection, for N$^{\varepsilon}$H of His (Figs.**12a** and **c**) and NAV (Figs.**12b** and **d**). This robustness is very large, especially with SQ selection (Figs.**12a** and **b**), with full-width at half maximum excitation bandwidth of S$_{FWHM} \approx 60$ and 40 kHz for His and NAV respectively, whereas it is smaller with DQ selection: S$_{FWHM} \approx 40$ and 30 kHz (Figs.**12c** and **d**). As previously noted, HMQC experiments using DANTE trains are less robust to offset than those using SLP, although DANTE benefits from a slight increase in overall on-resonance excitation efficiency [40]. It must be noted that the robustness to $^{14}$N$^{SQ}$ offset of $^1$H-{$^{14}$N} *D*-HMQC 2D experiment with SLP excitation (S$_{FWHM}$/$\nu_R$ $\approx 0.6$-1.0) is much larger than that of the corresponding method with double CP MAS transfer (S$_{FWHM}$/$\nu_R \approx 0.25$) (Fig.**4d-AM** in [31]). It has very recently been demonstrated that optimizing the carrier frequency is recommended when using SLP and this is even more the case for D$_1^K$ (Figs.**11** and **12** in [59]).

### 5. Conclusions

In this work, various previously proposed schemes for [14]N excitation in [1]H-{[14]N} HMQC experiments have been analyzed and compared. This analysis was performed with selection of either single- (SQ) or double- (DQ) quantum levels of [14]N nuclei. DQ selection leads to a smaller efficiency than SQ selection, without any resolution enhancement, and is thus not generally recommended with a modern speed controller and MAS probe.

Conventional hard pulses are not efficient and should be avoided. The same recommendation is given for the XiX (X inverse X) scheme, which has recently been proposed for DQ selection. The most efficient methods are the sideband selective long pulse (SLP) or the $D_1^K$ (DANTE) train of pulses. It has been shown here that SLP excitation is much more efficient than $D_1^K$ trains for large $C_Q$ values. As SLP is also more robust with respect to offset, rf-inhomogeneity and experimental setup, this scheme should therefore be preferred, and is particularly recommended for use by non-specialists, due to its simplicity. For weak or moderate $C_Q$ values, the choice between the two schemes depends on the rf and magnetic field strengths, the spinning speed, and the [14]N isotropic frequency range. When this range is very large, SLP, which is very robust with respect to offsets, should also be the sequence of choice. In the case of paramagnetic samples, the [14]N isotropic shift range may even be larger than $\nu_R$ leading to folded sidebands. In this case, two 2D experiments with two different spinning speeds should be acquired in order to discriminate the center-bands from the spinning sidebands [59].

It must be emphasized that [1]H-{[14]N} HMQC experiments with SQ selection, which require the cancellation of the first-order quadrupole dephasing in order to be well-resolved along the [14]N dimension, benefit from several recent technical developments. These include new MAS speed controllers that provide very stable spinning speeds, and new MAS probes with automatic magic angle adjustment with a level of precision that meets the so-called 'STMAS requirements'.

It must also be noted that very recently, a new variant of the $SR4_1^2$ recoupling sequence has been proposed for [1]H-{X} D-HMQC experiments at ultra-fast MAS. This variant replaces the basic $\pi$-pulses with $90_{-45}90_{45}90_{-45}$ composite $\pi$-pulses to decrease the effects of [1]H-[1]H interactions [60]. As a result, this recoupling quenches the spin diffusion and hence long-lived coherence lifetimes can be obtained, which facilitate the detection of long N-H distances with this non-dipolar truncated sequence.

Finally, it must be noted that the [1]H-{[14]N} D-HMQC experiment is complementary to the double CP scheme recently proposed [31]. Both methods experimentally require the same 'STMAS specifications' and ultra-fast sample spinning to minimize the losses and maximize the indirect spectral width. Compared with D-HMQC using SLP excitation, the double CP method exhibits a lower robustness to offset, but a better [14]N resolution when no [1]H decoupling is used. However, it should be pointed out that the [14]N resolution of D-HMQC can be enhanced with [1]H decoupling, and this allowed for the first detection of the cross-peak between aliphatic [1]H and [14]N ammonium site in Histidine [61]. It must also be reminded that two parameters have to be optimized with the double CP scheme: the contact times and the rf-fields to match the Hartmann-Hahn conditions [62,63], which are narrow at ultra-fast MAS [64,65]. On the contrary, there is no such rf-matching with D-HMQC and, moreover, with SLP excitation the [14]N rf-field is small and efficient over a broad range. To conclude, it ought also to be noted that the [1]H resolution can be enhanced by using weak continuous wave [14]N irradiation during $t_2$ [65].

**Acknowledgments**

Institut Chevreul (FR 2638), Ministère de l'Enseignement Supérieur et de la Recherche, Région Hauts-de-France and the European Union (FEDER/ERDF) are acknowledged for supporting and partially funding this work. Financial support from the IR-RMN-THC FR 3050 CNRS for conducting the research is gratefully acknowledged. Authors also thank ANR-17-ERC2-0022 (EOS) and ANR-18-CE08-0015-01 (ThinGlass). This project has also received funding from the European Union's Horizon 2020 research and innovation program under grant agreement no. 731019 (EUSMI). OL acknowledges financial support from Institut Universitaire de France (IUF). The computational resources were partially provided by the Polish Infrastructure for Supporting Computational Science in the European Research Space (PL-GRID). PP is grateful for funding from the MOBILITY PLUS program (grant no. DN/MOB/172/V/2017).

**Supplementary Information**: Figs.S1-S4.

# Evaluation of excitation schemes for indirect detection of $^{14}$N *via* solid-state HMQC NMR experiments

Andrew G.M. Rankin,[1]* Julien Trébosc,[1,2] Piotr Paluch,[1,3] Olivier Lafon,[1,4] Jean-Paul Amoureux[1,5]*

[1] Univ. Lille, CNRS, Centrale Lille, ENSCL, Univ. Artois, UMR 8181 – UCCS – Unit of Catalysis and Chemistry of Solids, F-59000 Lille, France
[2] Univ. Lille, CNRS-FR2638, Fédération Chevreul, F-59000 Lille, France
[3] Centre of Molecular and Macromolecular Studies, Polish Academy of Sciences, Sienkiewicza 112, PL-90363 Lodz, Poland.
[4] Institut Universitaire de France, 1 rue Descartes, F-75231 Paris Cedex 05, France.
[5] Bruker Biospin, 34 rue de l'industrie, F-67166 Wissembourg, France.

* Corresponding authors: jean-paul.amoureux@univ-lille.fr   andrew.rankin@univ-lille.fr

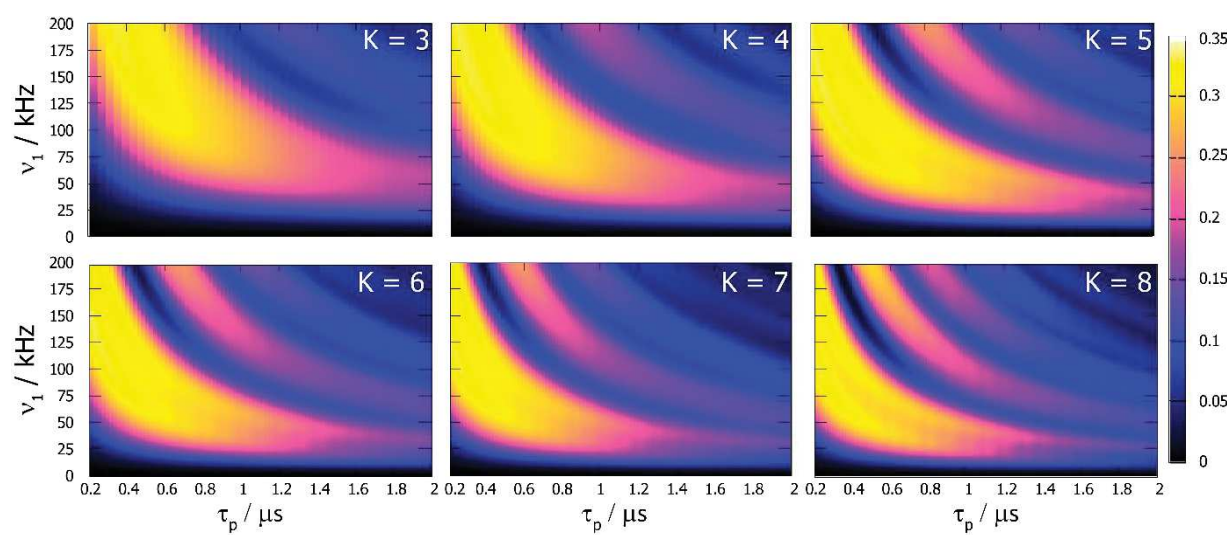

**Fig.S1.** 2D maps showing the simulated SQ efficiency of $D_1^k$ ($\nu_1$ *vs.* $\tau_p$) for a spin system corresponding to glycine, using K values between 3 and 8.

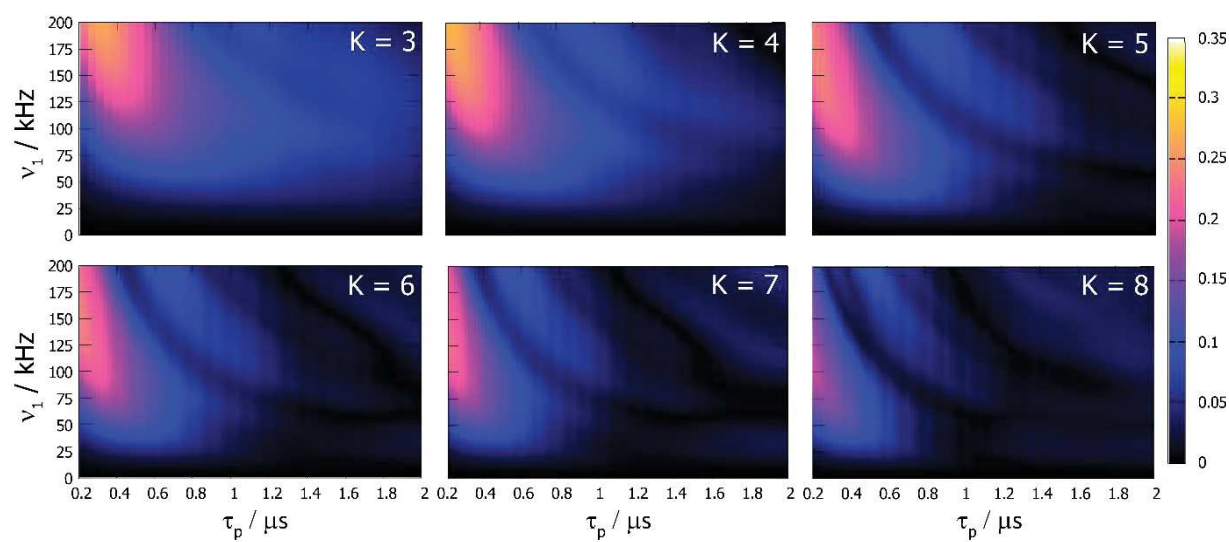

**Fig.S2.** 2D maps showing the simulated SQ efficiency of $D_1^k$ ($\nu_1$ *vs.* $\tau_p$) for a spin system corresponding to NAV, using K values between 3 and 8.

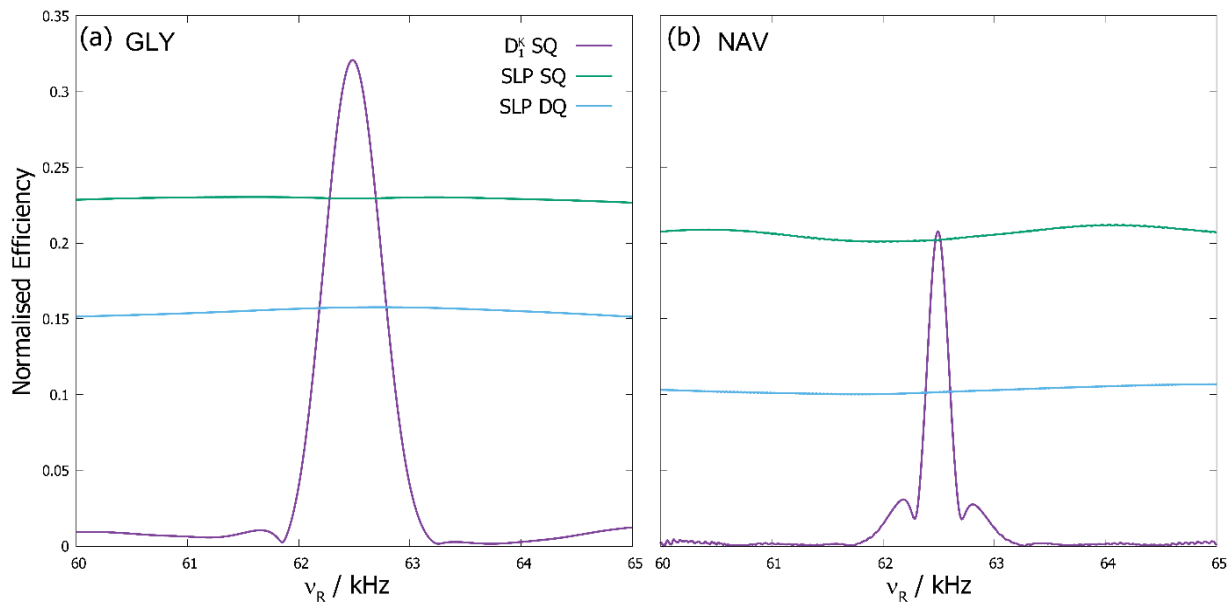

**Fig.S3.** 1D plots displaying the simulated efficiency for $D_1^k$ with SQ selection and SLP with SQ and DQ selection with respect to the MAS rate, $\nu_R$, for spin systems corresponding to (a) glycine and (b) NAV.

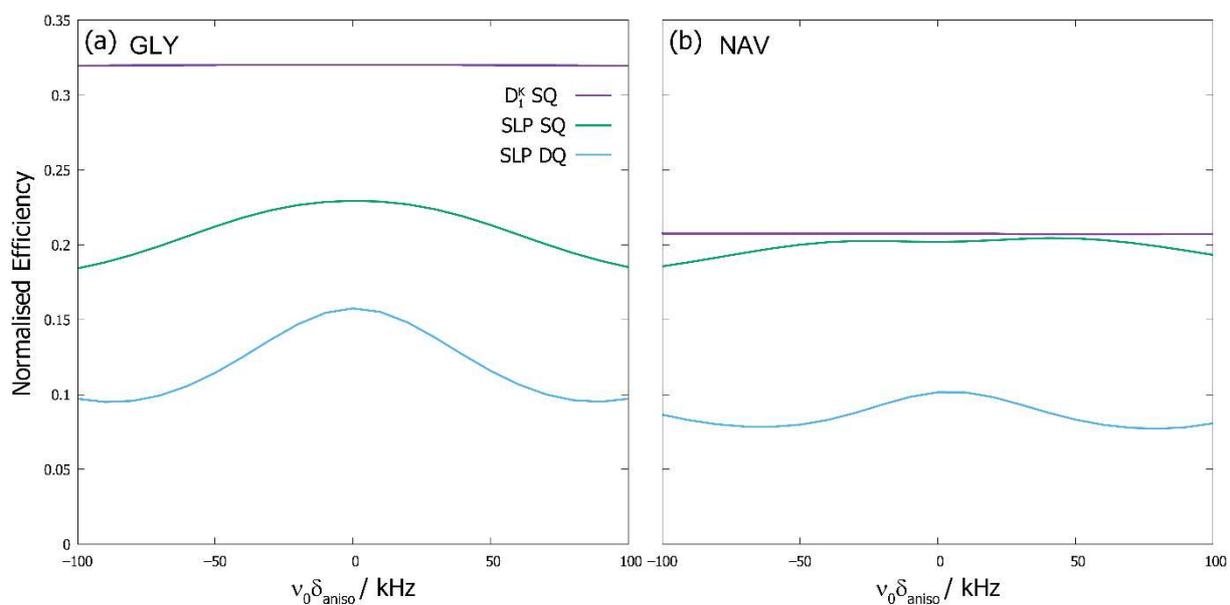

**Fig.S4.** 1D plots displaying the simulated efficiency for $D_1^k$ with SQ selection and SLP with SQ and DQ selection with respect to $^{14}N$ CSA ($\nu_0\delta_{aniso}$), for spin systems corresponding to (a) glycine and (b) NAV.